\documentclass[pdflatex,sn-mathphys-num]{sn-jnl}

\usepackage{graphicx}%
\usepackage{multirow}%
\usepackage{amsmath,amssymb,amsfonts}%
\usepackage{amsthm}%
\usepackage[title]{appendix}%
\usepackage{mathrsfs}%
\usepackage{xcolor}%
\usepackage{textcomp}%
\usepackage{manyfoot}%
\usepackage{booktabs}%
\usepackage{algorithm}%
\usepackage{algorithmicx}%
\usepackage{algpseudocode}%
\usepackage{listings}%

\theoremstyle{thmstyleone}%
\newtheorem{theorem}{Theorem}
\newtheorem{proposition}[theorem]{Proposition}%

\newtheorem{corollary}[theorem]{Corollary}%

\theoremstyle{thmstyletwo}%

\theoremstyle{thmstylethree}%
\newtheorem{definition}{Definition}%

\newtheorem{innercustomprop}{Proposition}
\newenvironment{customprop}[1]
  {\renewcommand\theinnercustomprop{#1}\innercustomprop}
  {\endinnercustomprop}

\newtheorem{innercustomcor}{Corollary}
\newenvironment{customcor}[1]
  {\renewcommand\theinnercustomcor{#1}\innercustomcor}
  {\endinnercustomcor}

\begin{document}

\title[Article Title]{Small singular regions of spacetime}


\author*[1,2]{\fnm{Franciszek} \sur{Cudek}}\email{franciszek.cudek@seh.ox.ac.uk}

\affil*[1]{\orgdiv{St Edmund Hall}, \orgname{University of Oxford}, \orgaddress{\street{Queen's Lane}, \city{Oxford}, \postcode{OX1 4AR}, \country{United Kingdom}}}
\affil*[2]{\orgdiv{Faculty of Philosophy}, \orgname{University of Oxford}, \orgaddress{\street{Woodstock Road}, \city{Oxford}, \postcode{OX2 6GG}, \country{United Kingdom}}}


\abstract{We prove that every open connected region of relativistic spacetime $(M,\textbf{g})$ that encloses a $b$-incomplete half-curve has an open connected subregion that encloses a $b$-incomplete half-curve and is also `small' in the following sense: it is the image, under the bundle projection map, of some open region in the (connected) orthonormal frame bundle $O^+M$ over that spacetime which is bounded, and whose closure is Cauchy incomplete, with respect to any `natural' distance function on $O^+M$. As a corollary, it follows that every $b$-incomplete half-curve can be covered by a sequence of singular regions which are images of a sequence of bounded subsets of $O^+M$ whose diameter, with respect to any `natural' distance function on $O^+M$, tends to zero. We discuss to what extent these results can be interpreted in favour of the claim that singular structure in classical general relativity is `localizable'.}




\maketitle

\bmhead{Acknowledgements}
Thanks to Silvester Borsboom, Jeremy Butterfield, Henrique Gomes, Klaas Landsman, James Read, and two anonymous referees for their comments and discussion of this material.

\bmhead{Funding}
Support for this work was provided by the AHRC, UK and the Clarendon Fund, Oxford.

\bmhead{Conflict of Interest Statement}
The author declares no conflicts of interest.

\bmhead{Data Availability Statement}
No data are associated with this article.

\newpage
\section{Introduction}\label{sec1}

In classical general relativity, a (relativistic) spacetime $(M, \textbf{g})$ is said to be {\em singular} just in case it contains curves that are incomplete in some appropriate sense. Different authors take different kinds of curves to witness singular structure of relativistic spacetimes: incomplete timelike or causal geodesics \cite{wald1984}; incomplete causal curves of bounded acceleration \cite{geroch1968b},\cite{olmo2018}; or $b$-incomplete curves \cite{hawking-ellis1973},\cite{clarke1993}. Here we will focus on spacetimes that are singular in the weakest of these three senses: namely, those which contain $b$-incomplete curves. 

Still, whatever kind of curves one designates to witness singular structure, the resulting definition of `singular spacetime' would arguably be insufficient to account for all the intuitions we have associated with singular structure in classical general relativity. Among other things, it seems plausible that in certain singular spacetimes, some regions of such spacetimes are singular, whereas others are not. For example: in the Kruskal--Schwarzschild spacetime, it seems true to say that Region I is not singular, whereas Region II is. To account for this, one cannot treat such proper regions as spacetimes on their own and then apply some general definition of `singular spacetime'; for any such spacetime would be extendible and therefore contain an incomplete timelike geodesic \cite[p. 8]{clarke1993} and count as singular on any plausible definition. This motivates the following generalisation of the definition of `singular spacetime' to arbitrary connected spacetime regions:
\begin{definition}
    For any relativistic spacetime $(M, \textbf{g})$, an open connected subset $U$ of $M$ is a \textit{singular region} of $(M, \textbf{g})$ iff there is an incomplete half-curve $\gamma: [0,a)\rightarrow M$ (for $0 < a \leq \infty$) of appropriate kind whose image is contained in $U$.
\end{definition}

This definition invites the following conceptual question: if singular structure in classical general relativity can be considered a property of spacetime regions, is it always possible to find, in any singular spacetime, a sufficiently `small' region of that spacetime that is singular? A positive answer to that question would have a desirable consequence that singular structure of spacetime would be, in some sense, localizable. Since we focus on spacetimes that contain $b$-incomplete curves, we will focus on $b$-incomplete singular regions: that is, regions that contain the images of $b$-incomplete half-curves. And we will show that for this definition of `singular region', any singular spacetime will have a singular region which is, in a way we will make precise, `small'.

Section 2 contains technical preliminaries. In Section 3, we pin down some aspects of the relationship between: (a) $b$-incompleteness of regions of a given spacetime, and (b) boundedness and Cauchy incompleteness of metric subspaces of the orthonormal bundle over that spacetime. More precisely, we prove the following result: 

\begin{customprop}{1}
    Let $(M, \textbf{g})$ be a relativistic spacetime, $U$ be an open subset of $M$, and $O^+M$ be the connected component of the orthonormal frame bundle over $(M, \textbf{g})$. Then, if there exists a curve $\gamma: [0,a) \rightarrow M$ with finite generalised affine length, and no endpoint, whose image is contained in $U$, then there exists a $b$-incomplete singular region $V \subseteq U$ and an open $\tilde{V} \subseteq O^+M$ such that: (i) $\tilde{V}$ is bounded, (ii) $\text{cl}(\tilde{V})$ is Cauchy incomplete (both (i) and (ii) understood with respect to \textit{any} `natural' distance function on $O^+M$), and (iii) $\pi[\tilde{V}] = V$.
\end{customprop}

We also prove the following corollary of Proposition \ref{prop3.1}:

\begin{customcor}{2}
    Let $(M, \textbf{g})$ be a relativistic spacetime, $U$ be an open subset of $M$, and $O^+M$ be the connected component of the orthonormal frame bundle over $(M, \textbf{g})$. Then, if there exists a curve $\gamma: [0,a) \rightarrow M$ with finite generalised affine length, no endpoint, and whose image is contained in $U$, and $\{t_n\}_{n\in\mathbb{N}} \rightarrow a$ as $n \rightarrow \infty$, then there exists a sequence $\{\tilde{V}_n\}_{n\in\mathbb{N}}$ of open subsets of $O^+M$ such that, for any natural distance function $d$ on $O^+M$: (i) each $\tilde{V}_n$ is bounded and $\text{cl}(\tilde{V}_n)$ is Cauchy incomplete, (ii) $V_n := \pi(\tilde{V}_n)$ is a singular region of $(M,\textbf{g})$ which contains the image of $\gamma|_{[t_n, a)}$, and (iii) the diameter of $\tilde{V}_n$ tends to zero as $n \rightarrow \infty$.
\end{customcor}

We also note that the `converse' relationship between Cauchy incompleteness and $b$-incompleteness---that any open connected subset of the (connected) orthonormal frame bundle whose closure is Cauchy incomplete with respect to any choice of the `natural' metric projects onto an open $b$-incomplete singular region of spacetime---is a relatively straightforward consequence of a theorem due to Hawking and Ellis \cite{hawking-ellis1973}, which extends and draws on the results of Schmidt \cite{schmidt1971},\cite{schmidt1973}.

Even though these results are of conceptual interest, we remark that the proofs involve little but straightforward metric space theory and rely on techniques and results that were available by the time Hawking and Ellis published their classical monograph \cite{hawking-ellis1973}. Also, it is worth stressing that Schmidt developed the notions of $b$-incompleteness and natural distance functions on the frame bundle to propose a new boundary construction for relativistic spacetimes and identify singularities with points on that boundary. Our approach, however, is manifestly different: we do \textit{not} propose any boundary construction, nor do we identify singularities with any points or regions. We use the concepts and techniques developed by Schmidt to extract a novel metric-invariant statement about the `localization' of singular behaviour in an arbitrary singular spacetime. Indeed, in Section 4, we briefly discuss to what extent our results might be interpreted in favour of the claim that singular structure is `localizable' in classical general relativity. 

\section{Preliminaries}\label{sec2}

First, we will introduce the notions of generalised affine parameter and $b$-incompleteness. Then, we will introduce the notion of `natural' Riemannian metrics on the orthonormal frame bundle over spacetime \cite{schmidt1971},\cite{hawking-ellis1973},\cite{dodson1978},\cite{clarke1993}.

A {\em (relativistic) spacetime} is a pair $(M, \textbf{g})$, where $M$ is a connected, Hausdorff, smooth $4$-manifold, and $\textbf{g}$ is a smooth Lorentz metric of signature $(3,1)$.\footnote{See \cite{wald1984} and \cite{oneill1983}. We use regular brackets for $n$-tuples of models and boldface font for tensors (as in \cite{hawking-ellis1973}).} We also assume that $(M, \textbf{g})$ is time- and space-orientable. This assumption is essential for Theorem \ref{thm3.3} and Proposition \ref{prop3.4}, but can be dropped for Proposition \ref{prop3.1} and Corollary \ref{cor3.2} (see n. 4 below).

Let $I$ be a half-open interval $[0, a)$ with $0 < a \leq \infty$. Given any $C^1$ curve $\gamma: I \rightarrow M$, we define the notion of a \textit{generalised affine parameter} $\lambda$ for $\gamma$ as follows. Given any basis $\{\textbf{e}_i\}$ for $T_{\gamma(0)}M$ (with $1 \leq i \leq 4$), extend this basis to tangent spaces on all of $\gamma[I]$ by parallel transport (so that $\{\textbf{e}_i|_{\gamma(t)}\}$ is a basis of $T_{\gamma(t)}M$). Then, for any $t \in I$, any vector $\textbf{X}|_{\gamma(t)}$ in $T_{\gamma(t)}M$ can be represented as $\textbf{X}|_{\gamma(t)} = \sum_{i}X^i(t)\textbf{e}_i|_{\gamma(t)}$. Now, given a curve $\gamma: I \rightarrow M$ with tangent vectors $\textbf{V}|_{\gamma(t)}$, the \textit{generalised affine parameter} $\lambda$ for $\gamma$, relative to the choice of basis $\{\textbf{e}_i|_{\gamma(0)}\}$ for $T_{\gamma(0)}M$, is given by:
    $$\lambda(t) = \int_0^t\left(\sum_{i}V^i(t')^2\right)^{1/2}dt',$$
    where $V^i(t')$ are components of $\textbf{V}|_{\gamma(t')}$ in the given basis.
Importantly, whether the length of a curve $\gamma$ as defined with respect to a given generalised affine parameter (known as `generalised affine length') is finite, does \textit{not} depend on the choice of bases \cite[p. 259]{hawking-ellis1973}. So, we speak of any curve as having finite, or infinite, generalised affine length without reliance on any particular choice of bases for the tangent spaces along the curve. This allows us to say, without any further specification, that a curve $\gamma: I \rightarrow M$ is \textit{$b$-incomplete} iff it has finite generalised affine length and no endpoint (that is, there is no point $p \in M$ such that for every neighbourhood $U$ of $p$ there exists $t \in I$ where $\gamma(t') \in U$ for all $t' > t$). We then say that a spacetime $(M, \textbf{g})$ is \textit{$b$-incomplete} iff it contains a $b$-incomplete curve.

Now, let us turn to some basic definitions from the theory of bundles. The \textit{frame bundle} $FM \xrightarrow{\pi} M$ over a  spacetime $(M, \textbf{g})$ is the principal fibre bundle with structure group $GL(4, \mathbb{R})$, where $FM$ is the set of all pairs of the form $(p, \{\textbf{e}_i|_p\})$, ranging over all $p \in M$ and all frames $\{\textbf{e}_i|_p\}$ at $p$.\footnote{For a general definition of a principal fibre bundle, see \cite[p. 50]{kobayashi-nomizu1963}. For some examples of physical interest, see \cite[pp. 221--24]{isham1999}. It is worth mentioning that the projection map for fibre bundles is a submersion, and therefore open \cite[p. 271]{lee2013}.} Any $g \in GL(4, \mathbb{R})$ acts freely on points in $FM$ from the right by its usual action on the frame. If $M$ is orientable, $FM$ has two connected components. From now on, we assume that an orientation on $M$ is chosen, and we restrict our attention to the positive connected component $F^+M$ of $FM$ and the structure group acting on it is the component of $GL(4, \mathbb{R})$ connected to the identity.

We say that the \textit{lift} $\overline{\gamma}$ of a curve $\gamma: [0, a) \rightarrow M$ at a point $u = (\gamma(0), \{\textbf{e}_i|_{\gamma(0)}\})$ in $F^+M$ is given by the curve $\overline{\gamma}: [0, a) \rightarrow F^+M$ such that: $\overline{\gamma}(0) = u$, $\pi[\overline{\gamma}(t)] = \gamma(t)$, and where the frame at any point $\overline{\gamma}(t)$, for $0 \leq t < a$, is obtained by parallel transport of $\{\textbf{e}_i|_{\gamma(0)}\}$ from $\gamma(0)$, along $\gamma$, to $\gamma(t)$. In this way, the Levi-Civita connection on $M$ also provides a connection one-form $\omega$ on $F^+M$ which, for each $u \in F^+M$, gives the map $\omega_u: T_uF^+M \rightarrow \mathfrak{gl}(4, \mathbb{R})$, whereas the frame bundle structure gives the canonical one-form $\theta$ which, for each $u \in F^+M$, gives the map $\theta_u: T_uF^+M \rightarrow \mathbb{R}^4$ \cite{kobayashi-nomizu1963},\cite{schmidt1971}.

Now, suppose $\langle\cdot,\cdot\rangle_{\mathfrak{gl}(4, \mathbb{R})}$ and $\langle\cdot,\cdot\rangle_{\mathbb{R}^4}$ are Euclidean inner products on $\mathfrak{gl}(4,\mathbb{{R}})$ and $\mathbb{R}^4$, respectively. Following Schmidt \cite{schmidt1971} and Marathe \cite{marathe1972}, we define a Riemannian metric $\textbf{h}$ on $F^+M$, relative to a particular choice of inner products, by its pointwise action on arbitrary vector fields $\textbf{X}$ and $\textbf{Y}$ on $F^+M$, as follows:
$$\textbf{h}(\textbf{X},\textbf{Y})|_u = \langle\omega_u(\textbf{X}|_u), \omega_u(\textbf{Y}|_u)\rangle_{\mathfrak{gl}(4,\mathbb{{R}})} + \langle\theta_u(\textbf{X}|_u),\theta_u(\textbf{Y}|_u)\rangle_{\mathbb{R}^4}$$

As with any Riemannian metric on a connected manifold, $\textbf{h}$ defines the canonical distance function $d: F^+M \times F^+M \rightarrow \mathbb{R}$ thereby turning $F^+M$ into a metric space (there is no canonical distance function definable from the metric when the manifold is not connected---hence our restriction to $F^+M$). We will refer to any thus-constructed Riemannian metric $\textbf{h}$ as a \textit{natural} metric on $F^+M$, and to the associated distance function $d$, definable from any such metric $\textbf{h}$, as a \textit{natural} distance function on $F^+M$. Such metrics are `natural' in the sense that they are derived from the Levi-Civita connection on $M$ that is uniquely specified by the metric tensor $\textbf{g}$ on $M$. Any two natural metrics $\textbf{h}_1$ and $\textbf{h}_2$, which differ over the choice of Euclidean inner products on $\mathfrak{gl}(4, \mathbb{R})$ and $\mathbb{R}^4$, give rise to uniformly equivalent distance functions $d_1$ and $d_2$: that is, there exist $a, b \in \mathbb{R}^+$ such that for any $u,w \in F^+M$, $ad_1(u,w) \leq d_2(u,w) \leq bd_1(u,w)$ \cite{schmidt1971},\cite{dodson1978}.

Now, suppose $H$ is a closed Lie subgroup of $GL(4,\mathbb{R})$ and $HM \xrightarrow{\pi_H} M$ is the principal fibre bundle over $M$ with the structure group $H$. If there is a smooth bundle morphism $\phi: HM \rightarrow FM$ that agrees with bundle projections and commutes with right actions, and the pullback of the connection on $FM$ induced by the Levi-Civita connection on $M$ by $\phi$ defines a connection on $HM$, the construction of the natural Riemannian metrics can be restricted to $HM \xrightarrow{\pi_H} M$ \cite{kobayashi-nomizu1963},\cite{friedrich1974}. In particular, for time- and space-orientable spacetimes, the component $SO^+(3,1)$ of the Lorentz group $O(3,1)$ connected to the identity is a closed Lie subgroup of $GL(4, \mathbb{R})$, and the connection on the (connected) orthonormal frame bundle $O^+M \xrightarrow{\pi_{O^+}} M$ over spacetime with structure group $SO^+(3,1)$ can be pulled back from that on $FM \xrightarrow{\pi} M$; so we can restrict our natural Riemannian metric to $O^+M$, and define associated distance functions in the usual way \cite{friedrich1974},\cite{dodson1978}.

\section{Main result}\label{sec3}
Now we prove the following result:

\begin{proposition}
\label{prop3.1}
    Let $(M, \textbf{g})$ be a relativistic spacetime, $U$ be an open subset of $M$, and $O^+M$ be the positive connected component of the orthonormal frame bundle over $(M, \textbf{g})$. Then, if there exists a curve $\gamma: [0,a) \rightarrow M$ with finite generalised affine length, and no endpoint, whose image is contained in $U$, then there exists a $b$-incomplete singular region $V \subseteq U$ and an open $\tilde{V} \subseteq O^+M$ such that: (i) $\tilde{V}$ is bounded, (ii) $\text{cl}(\tilde{V})$ is Cauchy incomplete (both (i) and (ii) understood with respect to \textit{any} natural distance function on $O^+M$), and (iii) $\pi[\tilde{V}] = V$.
\end{proposition}
\noindent \textit{Proof.} \\
Consider any curve $\gamma$ in $(M, \textbf{g})$, and the lift $\overline{\gamma}$ of $\gamma$ to $O^+M$ such that: (a) $\overline{\gamma}(0)$ passes through $u = (\gamma(0), \{\textbf{e}_i|_{\gamma(0)}\})$, and (b) $\overline{\gamma}(t)$ passes through $(\gamma(t), \{\textbf{e}_i|_{\gamma(t)}\})$ for $\{\textbf{e}_i|_{\gamma(t)}\}$ obtained by parallel transport of $\{\textbf{e}_i|_{\gamma(0)}\}$ along $\gamma$. Then, the arc length of $\overline{\gamma}$ from the initial point $\overline{\gamma}(0)$, as measured by any natural metric $\textbf{h}$, is equal to the generalised affine parameter of $\gamma(t)$ with respect to the basis $\{\textbf{e}_i|_{\gamma(t)}\}$. So, if $\gamma$ has finite generalised affine length, then $\overline{\gamma}$ has finite arc length with respect to any $\textbf{h}$. Now, assume $\gamma: [0,a) \rightarrow U$, for $U \subseteq M$, is a curve with finite generalised affine length and no endpoint in $M$. Take any point $u = (\gamma(0), \{\textbf{e}_i|_{\gamma(0)}\})$ in $O^+M$, and suppose that the generalised affine length of $\gamma$ with respect to the parallelly propagated basis $\{\textbf{e}_i|_{\gamma(0)}\}$ is $b$. (It follows that the arc length of the lift $\overline{\gamma}$ of $\gamma$ through $u$ is also equal to $b$.) Now, choose a particular natural metric $\textbf{h}_1$ on $O^+M$ with the associated natural distance function $d_1$, and consider, for any point $v \in \overline{\gamma}$, an open ball $B_\epsilon(v) \subseteq O^+M$ around $v$ of radius $\epsilon$. Then, the set $\tilde{W} := \bigcup_{v\in\overline{\gamma}}B_{\epsilon}(v)$ is open in $O^+M$ and bounded above, as a metric subspace of $(O^+M, d_1)$, by $b+2\epsilon$. It follows that $\tilde{V} := \pi^{-1}[U] \cap \tilde{W}$ is also open in $O^+M$ and bounded above by $b+2\epsilon$ in $(O^+M, d_1)$. Since any two natural distance functions are uniformly equivalent, these boundedness claims hold independently of any particular choice of $\textbf{h}_1$ (although the bounds might differ). Also, since the projection map in fibre bundles is open, $V := \pi[\tilde{V}]$ is an open subset of $U$; and it also contains $\gamma$, so it is singular in the $b$-incomplete sense.

Now, given any choice of the natural distance function $d$, assume, for reductio, that $\text{cl}(\tilde{V})$ is Cauchy complete with respect to $d$.\footnote{Strictly speaking this should be the restriction of $d$ to $\text{cl}(\tilde{V})$, but I omit that for notational hygiene.} Recall that $b$ is the generalised affine length of $\gamma$ with respect to the parallely propagated basis $\{\textbf{e}_i|_{\gamma(0)}\}$, and consider the sequence of points $\{x_n\}_{n \in \mathbb{N}^+}$ on $\overline{\gamma}$, where $x_n$ is the lift of the point in $\gamma$ with generalised affine length $b - (b/(2^n))$. This sequence is Cauchy: since for any $c \in \mathbb{R}^+$ and any $\epsilon > 0$, there exists $N \in \mathbb{N}^+$ such that $1/2^N < \epsilon/c$, we have it that for any $\epsilon > 0$, there exist $x_i, x_j$ with $j > i$ such that  $d(x_i, x_j) < b(1/2^i - 1/2^j) < \epsilon$. Since $\text{cl}(\tilde{V})$ is assumed to be Cauchy complete, it contains the limit $x$ of $\{x_{n}\}$. Now, consider any open neighbourhood $W$ of $\pi(x)$ in $M$. Then, $\pi^{-1}[W]$ will be an open neighbourhood of $x$, and will contain infinitely many elements of $\{x_{n}\}$, since it will contain some open ball centred at $x$. Thus, $W$ will contain infinitely many points of $\pi(\{x_n\})$, and since the generalised affine length of $\gamma$ is finite, there will be $t_0$ such that $W$ contains $\gamma(t)$ for any $t > t_0$. But then, $\pi(x)$ would be an endpoint of $\gamma$, contrary to our assumption. So, $\text{cl}(\tilde{V})$ is Cauchy incomplete. \qed
\\

Proposition \ref{prop3.1} also leads to the following corollary, which shows that, for any singular spacetime $(M, \textbf{g})$, one can always find a sequence of subsets in the orthonormal frame bundle $O^+M$, which project onto singular regions of $(M, \textbf{g})$, and whose diameter tends to zero regardless of the choice of the natural distance function on $O^+M$.

\begin{corollary}
\label{cor3.2}
    Let $(M, \textbf{g})$ be a relativistic spacetime, $U$ be an open subset of $M$, and $O^+M$ be the positive connected component of the orthonormal frame bundle over $(M, \textbf{g})$. Then, if there exists a curve $\gamma: [0,a) \rightarrow M$ with finite generalised affine length, no endpoint, and whose image is contained in $U$, and $\{t_n\}_{n\in\mathbb{N}}$ is a sequence in $[0,a)$ such that $t_n \rightarrow a$ as $n \rightarrow \infty$, then there exists a sequence $\{\tilde{V}_n\}_{n\in\mathbb{N}}$ of open subsets of $O^+M$ such that, for any natural distance function $d$ on $O^+M$: (i) each $\tilde{V}_n$ is bounded and $\text{cl}(\tilde{V}_n)$ is Cauchy incomplete, (ii) $V_n := \pi(\tilde{V}_n)$ is a singular region of $(M,\textbf{g})$ which contains the image of $\gamma|_{[t_n, a)}$, and (iii) the diameter of $\tilde{V}_n$ tends to zero as $n \rightarrow \infty$.
\end{corollary}
\noindent \textit{Proof.}\\
Choose a particular distance function $d_1$ on $O^+M$, and for each $t_n$ in $\{t_n\}$, suppose that the restriction $\gamma_{t_n} := \gamma|_{[t_n,a)}$ of $\gamma$ is now the half-curve that witnesses the $b$-incompleteness of $(M,\textbf{g})$, as in the statement of Proposition \ref{prop3.1}. Proceeding as in the proof of Proposition \ref{prop3.1}, construct an open subset $\tilde{V}_n$, but with an extra requirement that, for any $v \in \overline{\gamma_{t_n}}$, the open ball $B_{\epsilon_n}(v)$ has radius $\epsilon_n := \epsilon/n$. By Proposition \ref{prop3.1}, it follows that $\tilde{V}_n$ is bounded, $\text{cl}(\tilde{V}_n)$ is Cauchy incomplete, and $V_n := \pi(\tilde{V}_n)$ is a singular region containing the image of $\gamma_{t_n}$. Moreover, these facts hold independently of our choice of $d_1$. Also, the diameter of $\tilde{V}_n$ is bounded by $b_n+2\epsilon_n$ with respect to $d_1$, and it is clear that this tends to zero as $n \rightarrow \infty$. But since natural distance functions on $O^+M$ are uniformly equivalent, these diameters will tend to zero for any choice of such distance function.\footnote{Note that the proofs of Proposition \ref{prop3.1} and Corollary \ref{cor3.2} would work just as well for the $F^+M \xrightarrow{\pi} M$ bundle.}\qed
\\

One can also derive the `converse' result to Proposition \ref{prop3.1}, which infers $b$-incompleteness of spacetime regions from Cauchy incompleteness of appropriate bundle subspaces; but, in that case, there is no need to assume that the appropriate bundle subspace is bounded---hence scare quotes around `converse'. To do this, we will rely on an important theorem due to Hawking and Ellis \cite{hawking-ellis1973}, which extends and draws on the results of Schmidt \cite{schmidt1971},\cite{schmidt1973}, and is proved only for $O^+M$. It might be considered a partial Lorentzian surrogate of the Hopf--Rinow theorem:\footnote{The Hopf--Rinow theorem states that the following properties of a connected Riemannian manifold $(M, \textbf{h})$ are equivalent: (a) $(M, \textbf{h})$ is geodesically complete, (b) $(M, d_\textbf{h})$, as a metric space, is Cauchy complete, and (c) every closed and bounded subset in $(M, d_\textbf{h})$ is compact. This theorem cannot be generalised to the Lorentzian setting in any straightforward way. First, unlike in the Riemannian case, there is no natural distance function definable from the metric tensor that can turn the Lorentzian manifold into a metric space. Second, there are examples of compact Lorentzian manifolds, such as the Clifton--Pohl torus, that are geodesically incomplete \cite[p. 193]{oneill1983}. So, when we say that Theorem \ref{thm3.3} might be considered a \textit{partial} Lorentzian surrogate of the Hopf--Rinow theorem, all we mean is that it relates a certain kind of curve incompleteness of a manifold with Cauchy incompleteness of a closely related space (somewhat analogously to the properties (a) and (b) in Hopf--Rinow), even though it says nothing about the Lorenztian surrogate for the notion of a bounded subset.}
\begin{theorem}[Schmidt, Hawking and Ellis]
\label{thm3.3}
    A spacetime $(M, \textbf{g})$ is $b$-incomplete iff, for any natural distance function $d$ on $O^+M$, $(O^+M, d)$ is Cauchy incomplete.
\end{theorem}
\noindent \textit{Proof.} \\
See Propositions 8.3.1 and 8.3.2 in Hawking and Ellis \cite[pp. 278--82]{hawking-ellis1973}. \qed
\\

Then, the right-to-left direction of Theorem \ref{thm3.3} allows us to prove the following:
\begin{proposition}
\label{prop3.4}
    If $\tilde{V}$ is an open, connected subset of $O^+M$ such that, for some choice of the natural distance function $d$, $\text{cl}(\tilde{V})$ is Cauchy incomplete, then $\pi[\tilde{V}]$ is a $b$-incomplete singular region of $M$.
\end{proposition}
\noindent \textit{Proof.} \\
Suppose $\text{cl}(\tilde{V})$ is Cauchy incomplete with respect to the restriction of some natural distance function $d$. Now, let $\{x_n\}_{n \in \mathbb{N}^+}$ be a Cauchy sequence witnessing the incompleteness of $\text{cl}(\tilde{V})$. One can approximate this sequence by a Cauchy sequence $\{y_n\}$ in $\tilde{V}$ that also does not converge in $\text{cl}(\tilde{V})$: let $d_i := d(x_i, x_{i+1})$ and, for any $x_i$, consider an open ball $B_{d_i}(x_i)$ of radius $d_i$ around $x_i$. Since $\tilde{V}$ is open in $O^+M$ and thus a topological manifold, it has no isolated points which ensures that $B_{d_i}(x_i)$ will contain some $y_i$ from $\tilde{V}$. For any such choice of $y_i$ and $y_{i+1}$, the distance $d(y_i, y_{i+1})$ is bounded by $2d_i+d_{i+1}$, so $\{y_n\}$ is Cauchy and approximates $\{x_n\}$ since $d(x_n, y_n) \rightarrow 0$ as $n \rightarrow \infty$ (so it also does not converge in $\text{cl}(\tilde{V})$). Now, we can take $\{y_n\}$ as a witness of incompleteness of $\tilde{V}$ and apply right-to-left direction of Theorem \ref{thm3.3} for $\pi[\tilde{V}]$ treated as spacetime on its own (which it can be, since it is connected and open in $M$, as $\pi$ is an open map), to get a curve $\gamma: [0, \tau_{\text{fin}}) \rightarrow \pi[\tilde{V}]$ with finite generalised affine length and no endpoint in $\pi[\tilde{V}]$. But this curve will have no endpoint in $\text{cl}(\pi[\tilde{V}])$ either, for otherwise $\{y_k\}$ would converge in $\pi^{-1}[\text{cl}(\pi[\tilde{V}])]$. And in that case, it would also converge $\text{cl}(\tilde{V})$, as it is contained in $\tilde{V}$, contrary to what we have shown.\qed

\section{Further discussion}\label{sec4}
Let us now turn to the conceptual moral that one can draw from these results. First, note that it is a common desideratum for a successful analysis of singular structure in classical general relativity that it would vindicate the idea that singular structure is, in some sense, localizable \cite{earman1995},\cite{curiel2009}. A particularly forceful and straightforward way of ensuring localizability is to demonstrate that singular structure is instantiated by objects---singularities---that can be represented by well-defined points on the boundary of some mathematical space which is structurally apt to represent a possible physical spacetime described by classical general relativity \cite{geroch1968a},\cite{schmidt1971},\cite{geroch-et-al1972},\cite{scott-szekeres1994},\cite{flores-et-al2016}. As is well-known, however, such boundary constructions face serious obstacles: some of them fail to appropriately separate points on the boundary from those in spacetime's interior; others misclassify the intuitive cases of singular structure; and yet others presuppose fairly severe causal conditions that a spacetime must meet in order to have a well-defined boundary \cite{bosshard1976},\cite{johnson1977},\cite{geroch-et-al1982},\cite{curiel2009}.

Faced with the obstacles regarding boundary constructions, one might seek a weaker sense in which singular structure can be localizable. In particular, one might hope that singular behaviour can be `localized' in a suitably `small' region of spacetime. Then, one could point at such a `small' region and say `This is where singular structure is exhibited'. An immediate idea might be that a region is `small' just in case it is bounded; but, of course, there is no natural notion of a `bounded region' in the Lorentzian setting, since the Lorentzian metric does not induce a standard distance function. Neither is it appropriate to rely on relative compactness: excisions of points or closed sets from arbitrarily small open regions can result in both geodesic incompleteness and failure of relative compactness, as witnessed most clearly by Minkowski spacetime with an excised point.\footnote{Nor is it appropriate to rely on the {\em failure} of relative compactness of causal diamonds. First, this cannot happen in globally hyperbolic spacetimes, some of which are singular \cite[p. 207]{wald1984}. Second, there are geodesically complete (and so non-singular) spacetimes which have causal diamonds that are not relatively compact, such as universal anti de Sitter spacetime \cite{calabi-markus1962},\cite{friedrich1995}. Nor is it appropriate to rely on the technical notion of `$b$-boundedness', for any $b$-complete spacetime is $b$-bounded \cite[p. 292]{hawking-ellis1973},\cite[p. 448]{dodson1978}.}

An alternative notion of a `small' region of a relativistic spacetime is offered by Proposition \ref{prop3.1}.\footnote{Another alternative, which we do not discuss here, would be to use the metric-induced spacetime volume. That concept has been recently used to introduce the notion of `volume incomplete' spacetimes \cite{garcia-heveling2024}.} For that proposition says, roughly speaking, that any singular region of spacetime will have a singular subregion that can be thought of as the `shadow' of some bounded and incomplete region of the orthonormal frame bundle over that spacetime (where the latter region is bounded and incomplete according to \textit{any} natural way of specifying the metric in the bundle). One could say that we can now `localize' the singular structure by pointing to the `shadow' subregion and claim that this is where singular structure is localized. This does more justice to the intuition behind the concept of localizability, because the region we are pointing at is an image of a \textit{bounded} region in the bundle under a smooth surjective map. Intuitively, if the bounded region in the bundle is sufficiently `small', then its image under any smooth surjection is sufficiently `small' as well. Corollary \ref{cor3.2} makes the point even stronger. For it says, roughly, that for any singular spacetime, we can always find a sequence of singular regions in that spacetime which approach `zero size' as the $b$-incomplete curve they cover `approaches the singularity'.

We believe that these results make progress in our understanding of what it might mean for singular structure in classical general relativity to be `localizable', without the need to define singularities as boundary points of extended spacetimes. Still, we must add a qualification: even if there a sense in which the `shadow' regions are `small', and tend to `zero size', in a well-defined mathematical sense, it is uncertain to what extent these facts render singular regions `localizable' in a physically meaningful sense. That is because the physical significance of natural Riemannian metrics on the frame bundle remains opaque, and whereas the fact that they are derived purely from the physically meaningful Levi-Civita connection on the spacetime manifold signifies their naturalness, their wider theoretical and practical applicability remains underexplored.\footnote{It is also worth stressing that there is no guarantee that the `shadow' subregion $V := \pi[\tilde{V}]$ will generally be relatively compact---since, by Theorem \ref{thm3.3}, $b$-incompleteness of $M$ entails Cauchy incompleteness of $O^+M$, $O^+M$ does not have the Heine--Borel property; so we have no grounds to suppose that the closed and bounded region of the bundle that is the closure of the `shadow's' preimage would be compact (if that were the case, the relative compactness of $V$ would follow from $\pi$ being continuous and $M$ being Hausdorff).} We hope, however, that future work will shed light on this issue.

\backmatter


\end{document}